\documentclass{article}
\usepackage{float}
\usepackage{listings}
\usepackage{algorithm}
\usepackage[noend]{algorithmic}
\usepackage{amsmath, amsthm, amssymb, amsfonts}
\usepackage{graphicx}
\usepackage{color}
\usepackage{xy}
\usepackage{url}
\usepackage{colortbl}
\usepackage{multirow}
\usepackage{xspace}

\usepackage{amsmath}
\DeclareMathOperator*{\argmax}{argmax}

\newcommand{\sysname}{Seriema\xspace}

\begin{document}

\title{\sysname: RDMA-based Remote Invocation\\with a Case-Study on Monte-Carlo Tree Search\\{\small {\color{red} Under peer review since Aug/2021}}}

\author{\textbf{Hammurabi Mendes}, Bryce Wiedenbeck, Aidan O'Neill\\
\texttt{\{\textbf{hamendes}, brwiedenbeck, aioneill\}@davidson.edu}\\
Davidson College, NC, USA
}

\maketitle

\begin{abstract}
We introduce \sysname, a middleware that integrates RDMA-based remote invocation, asynchronous data transfer, NUMA-aware automatic management of registered memory, and message aggregation in idiomatic C++1x. 
\sysname supports the notion that remote invocation and asynchronous data transfer are complementary services that, when tightly-integrated, allow distributed data structures, overlay networks, and Cloud \& datacenter service applications to be expressed effectively and naturally, resembling sequential code. In order to evaluate the usability of \sysname, we implement a Monte-Carlo Tree Search (MCTS) \emph{application framework}, which runs distributed simulations given only a sequential problem specification. Micro-benchmarks show that \sysname provides remote invocations with low overhead, and that our MCTS application framework scales well up to the number of non-hyperthreaded CPU cores while simulating plays of the board game \emph{Hex}.
\end{abstract}

\section{Introduction}

InfiniBand (IB) is a low-latency, high-throughput communication standard that allows applications to bypass the OS and perform Remote Direct Memory Access (RDMA) to remote machines in the network. Recent standard revisions by Mellanox feature communication latency on the order of 1 $\mu$s in their latest iterations\cite{ib-latency}.

RDMA data transfer requires addressing several details related to memory management, such as mapping (\emph{registering}) physical memory into the IB device's hardware, manually managing lifetime \& recycling for send/receive buffers of registered memory, and possibly even application re-engineering in order to accommodate critical polling operations directly associated with latency and throughput. While libraries exist to abstract some of these concerns~\cite{infinity}, this work supports the notion that RDMA data transfer alone is not a service comprehensive enough to fully support a large class of distributed applications, such as distributed data structures, overlay networks, Cloud services, and datacenter services such as scheduling or consensus. Many of these applications transmit small, latency-sensitive messages that eventually trigger large-scale, throughput-sensitive data transfers. We see \emph{asynchronous remote invocation} performed over RDMA, in the form of RPCs (remote procedure calls) or their object-oriented counterpart RMI (remote method invocation), as complementary to \emph{one-sided} RDMA-based data transfer, where remote memory is written to directly without any awareness of a destination process.
 
We present a system called \sysname that provides RDMA-based remote invocation, NUMA-aware automatic management of registered memory, asynchronous data transfer, and message aggregation in idiomatic C++1x.
Remote invocation and asynchronous data transfer are tightly-integrated in \sysname. Some of our remote invocation primitives condition execution to completing previous, associated data transfers holding the function's data. In addition, remote invocation and asynchronous data transfer share a common notification infrastructure (Sec.~\ref{Sec-RemoteInvocation}) and have identical configuration options related to (NUMA-aware) memory-management. Our design approach targets \emph{unobtrusiveness}, which we define as allowing developers to simultaneously access multiple levels of abstraction in \sysname, choosing the interface that bests suits the application as it currently is. Hence, an application can use routines ranging among (a) a simple RDMA library that abstracts some performance-enabling programming techniques; (b) a more comprehensive RDMA solution, that also provides automatic memory management (allocation \& recycling) and automatic selective signaling (Sec.~\ref{Sec-SelectiveSignalingMemoryManagement}); (c) a distributed system \emph{fabric}, which in addition allows processes and threads to be placed and named, and also provides basic remote invocation; up to (d) a complete solution to asynchronous remote invocation using one-sided RDMA (see Sec.~\ref{Sec-BackgrondRelatedWork}), with optional local aggregation of remote calls. A single application could access \sysname concurrently in any of these levels.

As a usability experiment, we implemented a distributed Monte-Carlo Tree Search (MCTS) application \emph{framework}, a choice based on the wide applicability of MCTS and its challenges for parallelization (Sec.~\ref{Sec-BackgrondRelatedWork}). Our MCTS framework, a higher-level middleware itself, accepts MCTS problem specifications to be run in a distributed fashion, without any user-provided communication or MCTS logic code. Micro-benchmarks show that \sysname has low overhead remote invocations, and our proof-of-concept MCTS framework scales well up to the number of non-hyperthreaded CPU cores in our distributed testing environment while simulating plays of the board game \emph{Hex}.

The main challenges addressed by \sysname are not only those related with the mechanics of RDMA (abstracting memory registration \& management, thread synchronization for selective signaling, etc) but also those related to \emph{usability} aspects of asynchronous remote invocation, particularly when complemented with one-sided RDMA. For example, challenges in the usability domain include (i) managing flow control when using remote buffers in a flexible manner, giving a variety of options to applications, including local buffering that honors RDMA's ``no-copy'' principle, by carefully managing registered memory; (ii) allowing notification of remote invocation to be setup in different ways, including a notification when either the send buffer could be reused, or when the remote thread consumed the function; (iii) designing serialization protocols that allow internal components of \sysname to identify partial writes in serialized functions (or data associated with functions), and abstract those from the applications; (iv) requiring no communication setup between short-lived threads, and only simple, one-shot setup routines for long-lived threads that communicate with one-sided RDMA; (v) providing message aggregation for remote invocations,  honoring the ``no-copy'' principle of RDMA; (vi) implementing NUMA awareness, which is critical for RDMA performance~\cite{rdmaNUMA}, but not only making NUMA abstracted from applications (in buffer memory registration \& management, for example), but also optionally exposed to applications, in cases where performance depends critically on precise thread/buffer placement; (vii) providing options to handle incoming remote invocations, so calls could be handled by application threads directly, or by process level consumers; (viii) allowing calls to block, use hardware transactional memory (HTM), or employ fine remote-memory synchronization operations, so that the application can implement a highly-diverse set of synchronization patterns (e.g. delegation~\cite{delegation}, combining~\cite{combining}, among others). Items (i), (iv), and (v) are particularly aimed at \emph{irregular applications}, namely those with communication patterns highly sensitive to the problem's input. With a combination of RDMA-based asynchronous remote invocation and RDMA data transfers, we expect to handle a wide variety of distributed applications.
 
Our MCTS framework would require considerable coding effort without \sysname. RDMA memory management (registration, recycling, buffer management \& flow control for remote invocation, NUMA awareness, etc), wait-free synchronization for RDMA selective signaling, protocol design for function serialization and completion notifications, queue-pair sharing \& configuration, among others. MCTS-related code represents less than 10\% of the total codebase necessary to compile our MCTS application framework, the rest coming from high-level, high-performance routines abstracted by \sysname. After background and related work in Sec.~\ref{Sec-BackgrondRelatedWork}, architecture overview in Sec.~\ref{Sec-ArchitectureOverview}, and implementation details in Sec.~\ref{Sec-Implementation}, Sec.~\ref{Sec-Experiments} we evaluate both usability and performance aspects of \sysname with the lens of our MCTS framework.


\section{Background and Related Work}
\label{Sec-BackgrondRelatedWork}

RDMA has been well-used in distributed applications, notably in the database domain~\cite{rdmaDB1,rdmaDB2,rdmaKV1,rdmaKV2}. Yet, some aspects of system integration, particularly with NUMA, are still under evaluation~\cite{rdmaNUMA} (our performance observations in Sec.~\ref{Sec-Experiments-MCTS} align with this work).

FaRM~\cite{rdmaFarm} provides a rich interface, with remote memory allocation, transactions, and a message-passing service with one-sided RDMA. We have different design and implementation perspectives: (i) we do not create a single shared memory space across the system; (ii) our remote invocation service is richer, with automatic memory management (NUMA-aware) and message aggregation still being ``zero-copy'' (Sec.~\ref{Sec-RemoteInvocationSubsystem}); (iii) besides one-sided RDMA remote invocation, we have send-based services for short-lived threads; (iv) all system-level synchronization for shared queue pairs, including those related to selective signaling and memory management, is done with a wait-free~\cite{progress} lightweight protocol, essential for latency-sensitive applications. We do not provide transaction \emph{services} for remote memory like FaRM, Storm~\cite{rdmaStorm}, or LITE~\cite{rdmaLITE}, but rather \emph{tools} to support transactions via asynchronous remote invocation: distributed lock managers, hardware transactional memory helpers, synchronization \& atomic operation helpers, etc. As our invocations are by default handled directly by the receiving threads, in FIFO order, implementing process-local transactions largely reduces to shared-memory concurrent programming inside remote invocations. In addition, we think that different local and distributed data types might be subject to highly-diverse optimizations (delegation~\cite{delegation}, combining~\cite{combining}, etc), better handled at the application-level not at the systems-level.

In~\cite{beyondMPI}, authors propose a higher-level abstraction for remote writes called RDMO (``O'' for operation), each performing a series of non-blocking read/write/atomic operations on a remote data structure identified under a manager service. The authors note that this interface complements (and relies on) remote invocations, rather than substitutes them. In the other direction, our framework provides certain features that approach, but not quite implement their abstraction. As seen in Sec.~\ref{Sec-RemoteInvocationSubsystem}, users can register functions to be executed by the receiving \texttt{RDMAMessenger}s with runtime identifiers. Those functions can perform synchronized accesses to local data structures identified by an application-level table, as in~\cite{beyondMPI}.

Our biggest source of inspiration is Grappa~\cite{grappa}. Its interface is elegant, but the system imposes user-level threading to applications, forces remote objects to reside in a partitioned global address space (PGAS), does not allow explicit NUMA-aware remote memory allocation, and does not permit much RDMA performance tuning, as it relies on RDMA-based MPI for communication. We consider Grappa's ``global address'' (representing a machine/address pair) and ``RDMA aggregator'' (which provides message aggregation) abstractions powerful and elegant. Our RDMA aggregation philosophy differs from Grappa, as we allow applications to establish \emph{optionally} private channels between any $t_1$ and $t_2$, with its internal buffers allocated and controlled by the \emph{sender} thread, avoiding any inter-thread data dependency that might introduce latency -- the details are given in Sec.~\ref{Sec-RemoteInvocationSubsystem}.

\subsection{Monte-Carlo Tree Search}

Monte-Carlo Tree Search (MCTS) is a state-space search algorithm with numerous applications in AI. Originally proposed for planning in Markov decision processes \cite{AdaptiveMultistageSampling}, MCTS has now many scientific and industrial applications.

%

We will describe the core algorithm for MCTS as applied to the board game \emph{Hex}~\cite{Hex}, but variations and extensions abound. Hex is played on a rhombus-shaped board made of hexagonal cells. Players alternate coloring any empty cell, and a player wins if they connect their edges (p1: top--bottom; p2: left--right) with an unbroken path of their color.

MCTS explores possible moves in a game with a sequence of \emph{rollouts}, which explore and expand a search tree incrementally, focusing effort on the most promising moves. Each rollout consists of four phases: selection, expansion, evaluation, and backpropagation. The selection phase traverses the MCTS tree from the root node representing the current game state down to some ``frontier'' node: one with unexplored moves.
At each node $n$, a ``move'' $m$ is chosen among $n$'s children, according to a selection policy, which trades off exploration (trying new moves) with exploitation (gathering more information about the best moves). The standard policy, UCB, selects according to:
\[ \argmax_{m} \; V\!AL_{m} + C\sqrt{\frac{\ln \left( VIS_n \right)}{VIS_m}} \]
where $V\!AL_m$ is the value estimate for move $m$, $VIS_n$ is the number of visits to this node, $VIS_m$ is the number of times move $m$ has been tried here, and $C$ is a scaling constant. Note that this formula is increasing in $V\!AL_m$, promoting exploitation, and decreasing in $VIS_m$ promoting exploration. UCB derives from an optimal solution to this tradeoff in the multi-armed bandit setting \cite{UpperConfidenceBound}, but numerous alternatives exist.

When selection reaches the frontier we move to the expansion phase, where an unexplored move is chosen (typically at random, as in our implementation) and added to the tree.  Thus each rollout creates one new leaf node as a child of the frontier node. The evaluation phase determines a value estimate for the newly-expanded node. Most straightforwardly, this evaluation can be performed by simulating random moves to the end of the game, but hand-crafted or learned state-value functions can also be employed.
Finally, the backpropagation phase updates the value estimates for nodes on the root-frontier path traversed during selection to incorporate the leaf-value obtained by evaluation. For a game like Hex, the value of a node corresponds to the win-rate for the player taking a turn there, and can be calculated as $V\!AL_m = \frac{WINS_m}{VIS_m}$.

A large amount of research and engineering effort has gone into parallelizing both the general MCTS algorithm and various domain-specific applications. Early efforts \cite{RootLeafParallel} sought ways to decompose the problem using either root parallelization, which runs several separate MCTS instances, or leaf parallelization, which runs one instance with multiple workers for the evaluation phase. A third approach, \emph{tree parallelization} \cite{TreeParallel}, where multiple threads traverse a shared tree structure, initially underperformed due to locking overhead, but with better implementations has become central to large-scale MCTS systems like AlphaGo \cite{AlphaGo}.

Tree-parallel MCTS requires not only that threads avoid data races, but also that threads avoid duplicating work. Despite its name, MCTS generally makes deterministic choices during the selection phase, meaning that multiple threads traversing the search tree simultaneously will follow the same root-frontier path, which in the worst case causes tree-parallel MCTS to approximate leaf-parallel with extra overhead.
The standard technique for avoiding this duplication is known as a \emph{virtual loss}, where threads pessimistically update the value of moves they choose during selection and then fix those updates during backpropagation. In the case of a game with only win/lose outcomes, this can be achieved by incrementing $VIS_m$ during selection and $WINS_m$ during backpropagation so that other threads visiting the same node will see a lower win rate for move $m$ in the interim.

Scaling parallel MCTS beyond shared memory systems requires partitioning the search tree. A common approach to partitioning is to compute a Zobrist hash---a deterministic function of the board state---that can be used to distribute search tree nodes uniformly across compute nodes \cite{ZobristHashing}. This approach is demanding on the communication system, as each step in the traversal can require a remote invocation. Recent work proposes topology-aware mapping heuristics \cite{TopologyHeuristic}, and hybrid approaches that incorporate aspects of root-parallelization \cite{RootTreeParallelization} in order to reduce the number of remote calls. Our experiments show that with \sysname (using additional local-memory concurrency optimizations, such as ~\cite{delegation,combining}), even a tree parallelization approach can scale, with micro-benchmarks showing a very low remote invocation overhead.

\section{Architecture Overview}
\label{Sec-ArchitectureOverview}

\begin{figure}[!htb]
	\centering
	\includegraphics[width=\columnwidth]{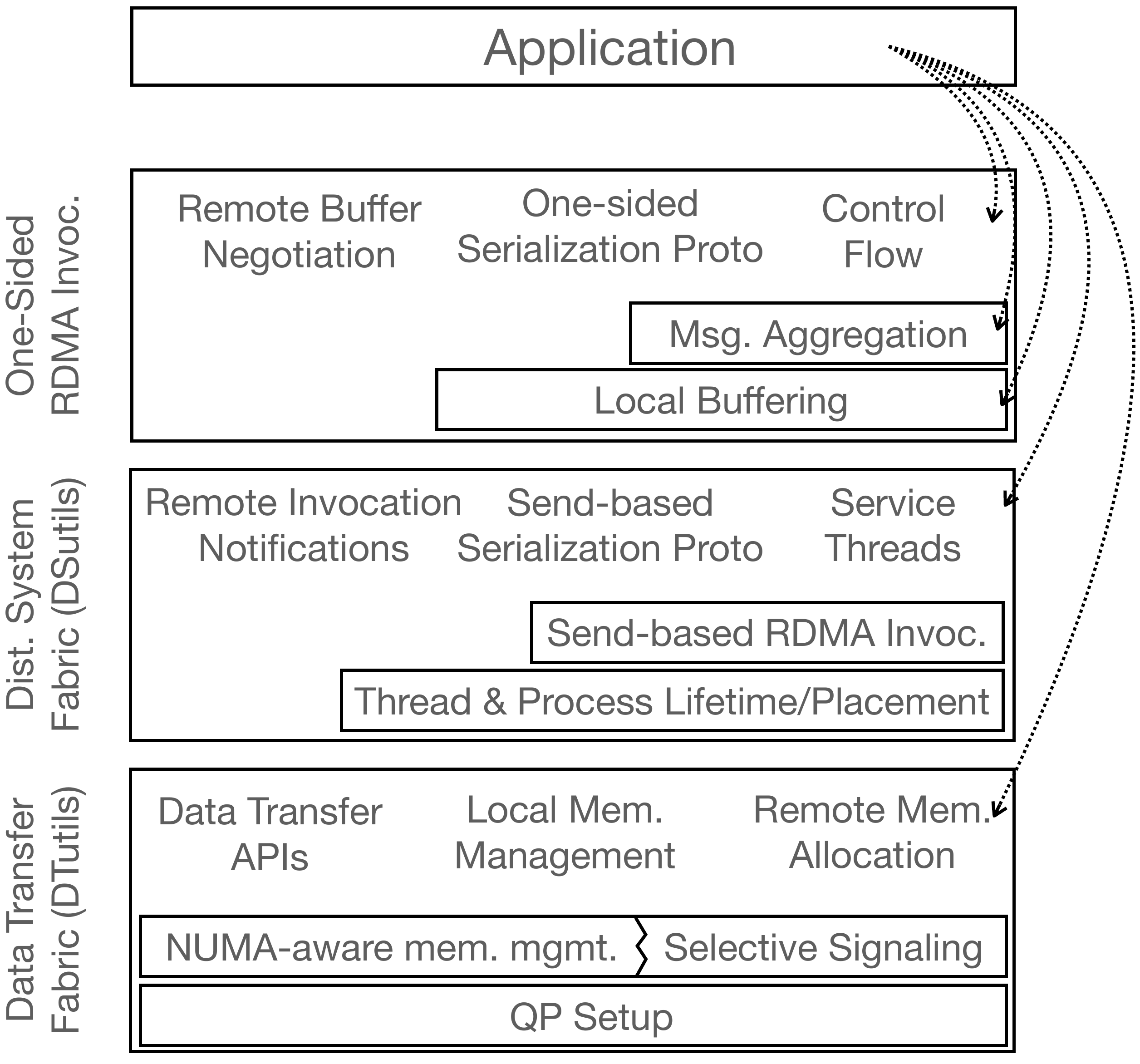}
	\caption{Architecture of \sysname.}
	\label{fig:overview}
\end{figure}

\sysname has three subsystem abstractions, all directly accessible by applications, as seen in Fig.~\ref{fig:overview}. In the lowest level, we have an RDMA \emph{Data Transfer Subsystem} called \texttt{DTutils}. It provides queue-pair state control, abstracts important performance-related techniques for RDMA, such as selective signaling and queue-pair sharing, and provides an efficient, automatic NUMA-aware memory management for RDMA (including fast memory registration and recycling).

In the intermediate level, we have a \emph{Distributed System Fabric}, called \texttt{DSfabric}, which initializes machines (e.g., creates and connects queue pairs according to users's specifications), initializes processes and long-lived application threads among machines, provides machine/process/thread identification, and provides basic remote invocation methods (at this level, only send-based).

In the highest level, we have a \emph{Remote Invocation Subsystem}, called \texttt{RIutils}. This module provides provides efficient asynchronous remote invocation with one-sided RDMA, with optional buffering and aggregation for remote calls (of course, both still honor RDMA's ``no-copy'' general principle).



We assume an application with a set of long-lived operating system \emph{processes}, running on a set of \emph{machines} connected via InfiniBand. Each machine contains a set of multiple NUMA \emph{zones}, and multiple InfiniBand \emph{endpoints} (a combination of device and port).

While we envision one process per (zone $\times$ endpoint) inside each machine, users have total control on process placement. Each process executes a set of \emph{worker threads}, consisting of user code, and a (possibly empty) set of \emph{service threads}, consisting of system support code. Worker threads can be short-lived or long-lived, and they might choose to perform service-thread roles if dedicated service threads are not used.

After establishing \emph{queue pairs} (QPs) between machines, relying on exchanging network interface ``local IDs'' (LID) and intra-device ``port'' numbers via an out-of-band channel, applications can use RDMA either with a send/recv interface -- which we call \emph{send-based} -- or with a read/write interface, called \emph{one-sided}. Send-based RDMA relies on the ``target'' machine registering receive buffers with the network interface, and on the ``source'' machine never exceeding the receive buffer sizes. Further, it also requires the ``target'' machine to consult whether receive buffers were consumed, typically via polling. One-sided RDMA performs reads/writes to memory at the ``target'' machine without awareness of the ``target'' application.

\subsection{Remote Invocation}
\label{Sec-RemoteInvocation}

The remote invocation primitives available both in send-based and one-sided RDMA are summarized in Table~\ref{Tab-RemoteCalls}. All primitives accept a function reference \texttt{func}, which is serialized into registered memory and executed remotely in the destination thread \texttt{dest}. C++ lambda captures allows users to include additional variables to the remote invocation context. All primitives accept a \texttt{Synchronizer} object called \texttt{sync}, which are semaphores that get decreased when \emph{either} (i) the function has been transmitted in the sender thread; or (ii) the function has been invoked in the receiving thread. The actual behavior of \texttt{sync} is specified via template parameters to the remote calls. When the semaphore resides in registered memory, and we operate under behaviour (ii), the (remote) semaphore notification is done using RDMA write.

The \texttt{call\_buffer} variants, besides including context variables via lambda captures, pass a buffer of unspecified type to the function executed remotely. The first variant of \texttt{call\_buffer} writes \texttt{orig\_buffer} (using RDMA write) into \texttt{dest\_buffer}, and only then calls \texttt{func} on the destination thread, passing a pointer to the fully-written \texttt{dest\_buffer} as an additional parameter. The second variant of \texttt{call\_buffer} calls a helper function on the destination thread that (a) copies \texttt{orig\_buffer} (using RDMA read) from the source thread into \texttt{dest\_buffer}, and only then calls \texttt{func} on the (local) destination thread, with the fully-written \texttt{dest\_buffer} as an additional parameter. Users can perform these steps either synchronously or in an asynchronous thread (controlled by a template parameter called \texttt{Async} in Tbl.~\ref{Tab-RemoteCalls}). For one-sided remote invocation (via \texttt{RDMAMessenger}s and \texttt{RDMAAggregator}s, described in Sec.~\ref{Sec-RemoteInvocationSubsystem}), we can pass the function and buffer of \texttt{call\_buffer} side-by-side in a single operation. The primitive \texttt{call\_return} accepts a pointer to a remote object that will be populated with the return of the invocation of \texttt{func} in the destination thread; and finally the \texttt{broadcast} and \texttt{broadcast\_buffer} primitives are respectively similar to \texttt{call} and \texttt{call\_buffer} but employ a broadcast tree in order to affect all threads.

\begin{table*}[t]
\centering
	\begin{tabular}{|p{0.56\linewidth} | p{0.40\linewidth}|}
	\hline
	Primitive & Effect \\
	\hline
	\hline
	\texttt{call}(dest, func, sync) & Thread \emph{dest} calls \texttt{func}(). \\
	\hline
	\texttt{call\_buffer}(dest, func, buffer, sync) & Thread \emph{dest} calls \texttt{func}(\emph{buffer}); buffer transmitted with the call. \\
	\hline
	\texttt{call\_buffer}(dest, func, orig\_buffer, dest\_buffer, sync) & Thread \emph{dest} calls \texttt{func}(\emph{dest\_buffer}); \emph{orig\_buffer} copied to \emph{dest\_buffer} via RDMA write. \\
	\hline
	\texttt{call\_buffer}(dest, func, orig\_buffer, sync) & Thread \emph{dest} calls \texttt{func}(\emph{buffer}); \emph{orig\_buffer} copied to \emph{buffer} via RDMA read. \\
	\hline
	\texttt{call\_return}(dest, func, orig, sync) & Thread \emph{dest} calls \emph{func}(); return value copied back into \emph{orig} via RDMA write. \\
	\hline
	\texttt{broadcast}(func, sync) & \texttt{call} on all threads using a broadcast tree. \\
	\hline
	\texttt{broadcast\_buffer}(dest, func, orig\_buffer, dest\_buffer, sync) & \texttt{call\_buffer} (1st variation) on all threads using a broadcast tree. \\
	\hline
	\texttt{broadcast\_buffer}(dest, func, orig\_buffer, sync) & \texttt{call\_buffer} (2nd variation) on all threads using a broadcast tree. \\
	\hline
	\end{tabular}
	\caption{Remote invocation primitives provided by \texttt{DSComm} (send-based) and \texttt{RDMAMessenger}/\texttt{RDMAAggregator} (one-sided). The \texttt{sync} parameter controls send/recv notification. Template parameters are discussed in the text.}
	\label{Tab-RemoteCalls}
\end{table*}

\subsection{Data Transfer}
\label{Sec-DataTransfer}

Data transfer implies and necessitates memory allocation, so this responsibility is part of our lowest-level \texttt{DTutils}. We allow applications to allocate memory locally or remotely. All allocations perform memory \emph{registration}, which is mapping its physical location into the IB interface's translation hardware. Registration is required by RDMA for most memory, except segments small enough to fit inside hardware interface buffers. \texttt{DTutils} internally implements techniques for fast memory registration~\cite{rdma-reg,rdma-reg2}, and provides a variety of memory allocators with different semantics (linear, circular, best-fit, etc). Details are discussed on Sec.~\ref{Sec-NUMA-MemoryRegistration}. 

We also expose to applications functions for \emph{remote memory allocation}, and allow users to constrain remote memory residence to a specific NUMA zone, named directly or indirectly by a (long-lived) application thread's name. Remotely-allocated memory is identified by a handle class called \texttt{RemoteMemoryLocator}.


\section{Implementation}
\label{Sec-Implementation}

Given an overview of \sysname, we now describe some implementation details that merit more careful presentation.



\subsection{Selective Signaling \& Memory Management}
\label{Sec-SelectiveSignalingMemoryManagement}

By default, work requests submitted to IB are \emph{signaled}, meaning that they will generate an RDMA completion queue entries upon finishing. Most commonly, however, we are not interested in completion events for every operation, and can thus reduce the overheard of generation and polling of completion events by making most operations \emph{unsignaled}. Each IB device has a maximum number of outstanding unsignaled transfers before they overflow (call it $u_\textrm{max}$), but when signaled operations complete, we are sure that all unsignaled operations that preceded it have also been completed. Hence, for all $k \bmod u_\textrm{max} = 0$, we can make only the $k$-th operations signaled, avoiding overflow and minimizing overhead.

With multiple threads, it is not enough to define an atomic counter $k$ for each operation within an RDMA QP, and poll its completion queue only when $k = 0 \bmod u_\textrm{max}$. In the time between the one where $k$ becomes $0 \bmod u_\textrm{max}$ and the one in which we flush the completion queue, other threads may perform operations and overflow the device. The solution is to define an additional atomic counter $k'$, indicating the number of times the queue pair's completion queue was flushed, called the \emph{flush number}. As long as $\lfloor k/u_\textrm{max} \rfloor > k'$, \emph{all threads} perform signaled operations, flush the QP's completion queue, and update $k'$ to the right flush number. The key feature, however, is that the flush number $k'$ not only avoids overflowing the device with unsignaled operations, but also helps recycle registered memory every $u_\textrm{max}$ operations.

We tag all \texttt{RDMAMemory} handles used with an \texttt{IBTransmitter} with its associated QP's identifier and current flush number just \emph{after} an operation is performed. When the \texttt{IBTransmitter}'s flush number increases, we know that all unsignaled operations tagged with smaller flush numbers are completed, and thus multiple memory regions are instantly freed with a single atomic update on memory. Higher abstractions for memory management automatically circulate buffers according to such tags (Sec.~\ref{Sec-NUMA-MemoryRegistration}). Our selective signaling / memory recycling protocol is wait-free~\cite{progress}, meaning that threads never block another thread's progress when they have to update flush numbers. We consider this property \emph{essential} for any low-latency system or service, particularly in latency-sensitive applications where the input might introduce high load imbalance and long data dependencies between threads.

\subsection{NUMA-Awareness and Memory Registration}
\label{Sec-NUMA-MemoryRegistration}


\texttt{DTutils} provides a variety of memory allocators to its users. In the most fundamental level, a set of concurrent \texttt{RDMAAllocator}s, one per NUMA zone, is responsible for retrieving NUMA-local memory in chunks of 1 GB superpages, registering the whole segment at allocation time. Registering superpages reduces the overhead of mapping physical memory into the device's memory management hardware multiple times~\cite{rdma-reg,rdma-reg2}. \emph{RDMAAllocator}s are still accessible directly by the application, from inside their respective machine, or even remotely. \texttt{RDMAAllocator}s also allow users to consume smaller chunks of 2MB superpages within each 1GB superpage.

In addition to \texttt{RDMAAllocator}, users have access to \emph{circular allocators}, \emph{linear-circular allocators}, and \emph{general allocators}.

A circular allocator subdivides 2MB chunks from \texttt{RDMAAllocator} into a circular buffer of \texttt{RDMAMemory} units. We move across units as long as they are free in the QPs in which they were last used, according to the protocol described in Sec.~\ref{Sec-SelectiveSignalingMemoryManagement}). In a circular buffer, note that we would be unable to move to a unit $a$ (free in a QP $q_a$) if we have some previous unit $b$ not free in a QP $q_b$. In order to avoid introducing latency because some QPs move slower than others, we allow latency-critical applications to create one circular allocator per \texttt{IBTransmitter} per thread -- an application-controlled tradeoff between latency and memory consumption. In circular allocators, the unit size is user-configured, and the total buffer size grows linearly or exponentially, up to a maximum limit, all user-configured.


The linear-circular allocator operates exactly like the circular allocators, but allows users to obtain \emph{segments} inside units, avoiding internal fragmentation. Finally, our general allocator obtains 1GB or 2MB chunks from \texttt{RDMAAllocator} and uses a traditional best-fit/first-fit strategies to allocate memory for applications. Chunks larger than 1GB are of course supported, but the superpage optimization described previously cannot be guaranteed for such large allocations.

\subsection{Send-Based RDMA}
\label{Sec-SendBased}

Recall that all of our invocation primitives are available in send-based or one-sided RDMA. In principle, send-based invocation seems inadequate, because (i) send-based invocations are limited by the size of the memory units posted by the service thread running on the receiver machine; and (ii) send-based RDMA has lower throughput than one-sided RDMA, because each call generates completion events on the service thread running on the receiver machine; and (iii) the handles for the buffers containing remote calls need to be further distributed to worker threads on the receiver machine.

Despite the shortcomings of send-based RDMA, particularly the overhead of generating and polling completion events, send-based calls do not require any \emph{a priori} negotiation of receive buffers between sender and receiver threads, a property that we call \emph{no-setup}. This is very convenient to implement higher-level services -- for example, our one-sided invocation services (\texttt{RDMAMessenger}s and \texttt{RDMAAggregator}) themselves use no-setup calls to allocate remote memory buffers across the system, and once this setup is performed, these buffers are used for one-sided RDMA operations. Given that flexibility, we allow applications to launch \emph{service threads}, which handle send-based remote invocation, and calls directed to and handled by service threads are denoted \emph{service calls}. We also allow service threads to receive send-based remote invocation directed at worker threads (not to the service thread itself), populating receive queues each associated with worker threads.

Both send-based and one-sided remote invocation require identification of the function to be executed remotely, which could be as simple as the function address -- when address-space layout randomization (ASLR) is disabled for processes\footnote{Generating non-position-independent executables also forces disabling ASLR.}. While we support this mechanism, we also allow users to (i) register their functions with application-level identifiers; or (ii) generate identifiers at compile time, because every lambda function \texttt{F} invoked remotely has a helper class \texttt{FunctionWrapper<F>}, and its type ID can serve as function identifier.

\subsection{The Remote Invocation Subsystem}
\label{Sec-RemoteInvocationSubsystem}

Our highest-level subsystem \texttt{RIutils} provides one-sided remote invocation and message aggregation. Our core abstraction for one-sided remote invocation is the \texttt{RDMAMessenger} class. It is created on a ``sender'' thread $s$ with a ``destination'' thread $d$ as a constructor parameter. Two \texttt{RDMAMessengers} with symmetrical sender/destination IDs are themselves called \emph{symmetrical}, and synchronize with each other, forming two channels for bidirectional communication. Each channel has its own set of memory buffers. For the sake of presentation, we will consider a single direction, calling one side \emph{sending} and the other side \emph{receiving}.

\subsubsection{Sender-Controlled Buffer Management}

Upon the first invocation (or a call to \texttt{setup()}) in the sending \texttt{RDMAMessenger}, the sending \emph{thread} performs a no-setup service call on the destination \emph{machine} in order to obtain $k$ handles of registered memory. These handles point to memory resident in the NUMA zone of the receiving \emph{thread}, and are represented by \texttt{RemoteMemoryLocator} objects. We call the remote memory buffer, which will subsequently receive serialized calls, a \emph{chunk}. As chunk allocations are done with no-setup service calls, there is no involvement of the receiving thread, reducing latency and avoiding deadlocks that would be possible otherwise. The sending \texttt{RDMAMessenger} stores handles for remote memory, and the receiving \texttt{RDMAMessenger} is notified of chunks operated remotely by the sending thread.

Users configure a parameter $c$, the number of handles obtained (or chunks allocated) per service call transaction, and a parameter $c_\textrm{max}$, the maximum number of obtained chunks (or allocated chunks). Our default settings have $c = 2$ and $c_\textrm{max} = 16$. Each handle points to a contiguous section of eight 2MB superpages in the receiving machine by default (the size user-configurable).

The no-setup service calls performing chunk allocation will: (i) allocate memory in the NUMA zone of the receiving thread; (ii) write the sending thread information on a header located inside the first allocated chunk; (iii) insert a handle for each allocated chunk into a concurrent, wait-free map ``owned'' by the receiving thread, where the key is the sending thread ID (the map is called \emph{incoming memory map}); and finally (iv) returns the chunk handle for the sending thread. Once the sending thread has obtained chunks, it may start serializing calls into them via RDMA write; once the receiving thread polls for messages in its receiving \texttt{RDMAMessenger}, it will note a new chunk associated with the sender thread in its incoming memory map, which will be extracted and incorporated into a circular buffer maintained by the receiving \texttt{RDMAMessenger}. The sending \texttt{RDMAMessenger} maintains a circular buffer of remote handles, and the receiving \texttt{RDMAMessenger} maintains a circular buffer of chunks. These symmetric circular buffers have to be kept ``symmetric'' even when new chunks are allocated (and new handles are obtained). We explain how this is done below.

Each chunk contains a \texttt{Header}, with two sub-structures called \texttt{Producer} and \texttt{Consumer}. The header is physically located in the receiving thread, and the \texttt{Consumer} section is modified only with local operations by the receiving thread; the \texttt{Producer} section is modified only with remote operations by the sending thread. The \texttt{Producer} header contains a monotonic, global offset of the first and last serialized calls within the chunk, updated by the sending thread when requested by users, or when a chunk is filled.

The destination thread updates the consumed offset in the chunk header via a local operation. At this point, the sending thread could technically obtain, in \emph{pulling} fashion, the last consumed offset by performing an RDMA read operation in the chunk header. This however introduces unnecessary waiting on the sender thread, so we implement a pushing mechanism as well. When the \emph{first} chunk is allocated on the receiving thread, information about the sending thread -- particularly, the memory location of the sending thread's \texttt{RDMAMessenger} -- is passed within the chunk header. Hence, the receiving thread notifies its last consumed offset in a \emph{pushing} fashion through an RDMA write operation to a variable inside the sending thread's \texttt{RDMAMessenger}, but only does that infrequently: this write happens only when a chunk is completely consumed. So now the sending thread can decide whether it can move forward within its circular buffer as the buffer gets fully consumed, without polling the receiver thread continuously.

If the sending thread cannot move forward within its circular buffer, it can allocate more chunks in the receiving machine up to $c_\mathrm{max}$, obtain the handles, and grow the handle circular buffer in the sending \texttt{RDMAMessenger}. The receiving thread will eventually notice the newly allocated chunks, and also incorporate them into the chunk circular buffer in the receiving \texttt{RDMAMessenger}. To ensure the the circular buffers remain perfectly symmetrical, whenever the sending thread finishes filling a chunk, it writes a special bit inside the \texttt{Producer} header of the associated chunk indicating the position where newly allocated chunks should be added in the receiving \texttt{RDMAMessenger}. This hints to the receiving thread to look into its incoming memory map in order to obtain new chunks, which have been allocated by the sending thread in order to expand the circular buffer.

Aside from this, note that the \emph{only} actual information about chunk consumption needed by the sending thread to decide whether it can rotate among its circular buffer of handles is a value written by the receiving thread in a pushing fashion. Hence, no remote invocations (and incurred latency) are needed at these particularly critical moments in the sending thread. If more chunks are needed, the protocol described in the previous paragraph allocates more memory without any involvement of the receiving thread (and incurred latency). If the sending thread reaches the limit of chunk allocations, remote invocations in the sending thread fail (that is, return false), until consumption picks up.

For convenience, we make available a higher-level class, called \texttt{RDMAMessengerGlobal}, that creates on demand the appropriate \texttt{RDMAMessenger}s at the sending thread upon the first communication attempt with any given destination thread.

\subsubsection{Aggregation \& Zero-Copy}

An \texttt{RDMAAggregator} is a class that operates on top of \texttt{RDMAMessenger}s and batches multiple remote calls into a single remote data transfer. We support two modes of operation: (i) a ``traditional'' approach, where remote invocations are batched locally, and flushed via \texttt{RDMAMessenger} to the destination thread at 4K intervals (size is user-configurable); or (ii) an ``overflow'' approach, where we perform local aggregation only when an underlying \texttt{RDMAMessenger} cannot make progress by its own. This occurs, for instance, when the sending thread is waiting for the destination thread to consume serialized calls and the number of allocated chunks has already reached $c_\mathrm{max}$. In the overflow approach, we denote the invocations batched locally in the sending thread as \emph{exceeding}. Users can specify memory consumption limits for exceeding invocations before invocations fail at the application level. Note that we honor RDMA's ``zero-copy'' principle by having exceeding invocations be serialized in registered memory, later transferred remotely -- noting lambda serialization into registered memory would need to happen regardless of aggregation.


\section{Experiments}
\label{Sec-Experiments}

We now describe our evaluation of: (i) the \texttt{DTutils} subsystem, via micro-benchmarks (Sec.~\ref{Sec-Experiments-DTutils}); (ii) the \texttt{DSfabric} and \texttt{RIutils}  subsystems, by assessing the cost of send-based and one-sided remote method invocation, as compared to the basic data transmission from \texttt{DTutils} (Sec.~\ref{Sec-Experiments-Serialization}); and (iii) the MCTS distributed framework implemented using our system (Sec.~\ref{Sec-Experiments-MCTS}). We have 4 machines, each with one dual-port QSFP Mellanox ConnectX-3 Pro VPI FDR InfiniBand switch, connected with the host by x8 PCI express 3.0. Each port on each device is connected to a Mellanox SX6012 X-2 FDR InfiniBand switch. The MTU of our interfaces is set to 4096. Each machine is a dual-socket Intel Xeon E5-2620v4, with 8 cores each running at 2.1GHz (varies between 2.0 and 2.5 with TurboBoosting), and 128GB of memory. Each core can run two hyperthreads (HT), giving a total of 16 cores (non-HT) or 32 cores (HT) per machine. The tool \texttt{numactl --hardware} reports relative intra-node distances of 10 and inter-node distances of 21 (no units). We use Open MPI 2.1.1 and compile with \texttt{gcc -O3 -std=c++17}.


\subsection{Microbenchmarks on \texttt{DTutils}}
\label{Sec-Experiments-DTutils}

\begin{figure*}[!htb]
	\centering
	\includegraphics[width=0.9\textwidth]{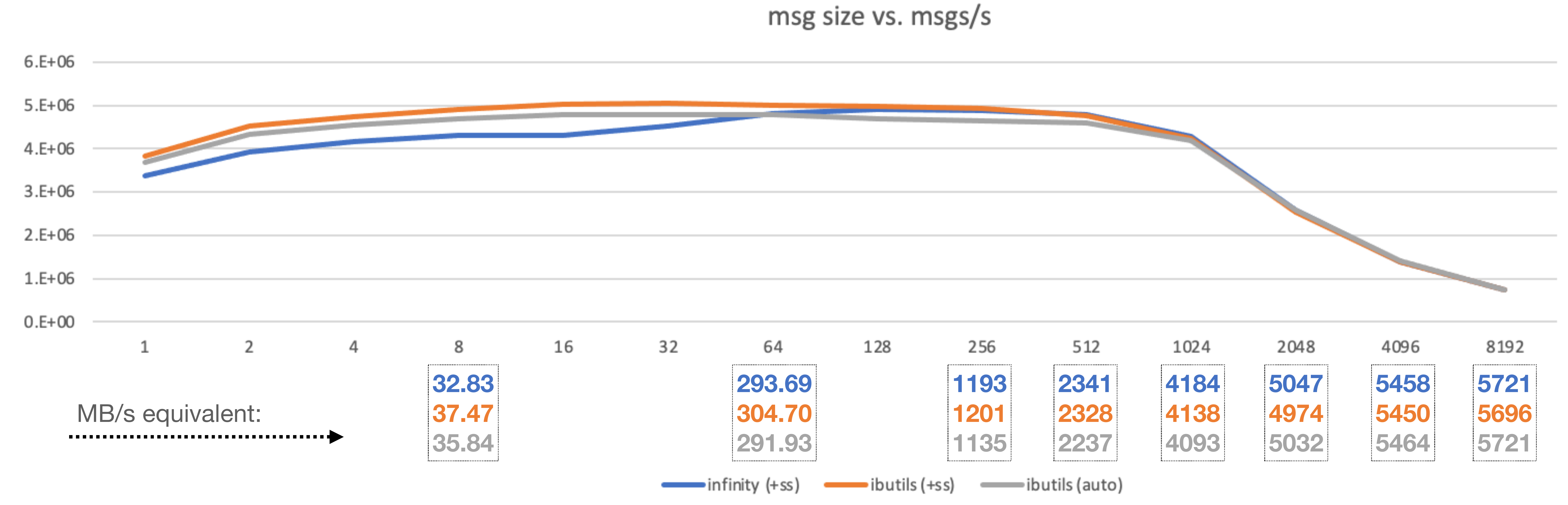}
	\caption{Message sizes in B vs. throughput in msg/s. Equivalent MB/s annotated under the $x$ axis.}
	\label{fig:DTutils_performance}
\end{figure*}

We first evaluate the \texttt{DTutils} subsystem, recalling it is also a \emph{detachable library}: applications can use it directly and separately from the other abstractions of \sysname. On Fig.~\ref{fig:DTutils_performance}, \texttt{infinity (+ss)} refers to the IB library from~\cite{rdmaDB2}, measured here for the sake of comparison. We implement selective signaling manually with that library. \texttt{DTutils (+ss)} refers to our own \texttt{DTutils}, accessing it through \texttt{IBQueuePair} directly, thus we also implement selective signaling manually in this case. Finally, \texttt{DTutils (auto)} refers to using our own \texttt{DTutils}, accessing it through \texttt{IBTransmitter}, thus selective signaling and circular buffer memory management are automatic, following the protocol of Sec.~\ref{Sec-SelectiveSignalingMemoryManagement}. 

Figure~\ref{fig:DTutils_performance} (p.~\pageref{fig:DTutils_performance}) shows, for each setting, the number of messages per second with varying message sizes, representing an average of three tests, with each test performing 128K transmissions in total. We provide the equivalent MB/s for some data points, as they will be useful in further analyses ahead.
The graph suggests that we have lower per-message overhead, which affects mostly small-sized messages. It also suggests that our automatic features introduce little overhead, as we compare \texttt{DTutils (+ss)} with \texttt{DTutils (auto)}. Note that our automatic selective signaling \emph{also} eliminates around 5 lines of C/C++ \emph{per remote invocation} in the user code related to selective signaling, and also takes care of memory management for the transmission circular buffer. A \emph{single line} of code is enough to provide remote invocation with both features enabled automatically when using \texttt{IBTransmitter}s.

\subsection{Remote Invocation Performance}
\label{Sec-Experiments-Serialization}

Table~\ref{tbl:call_performance} shows the throughput of \texttt{call} under: (i) send-based RDMA, provided by \texttt{DSfabric} (labeled \texttt{send}); one-sided RDMA, provided by \texttt{RDMAMessenger}s in \texttt{RIutils} (labeled \texttt{write}); (iii) overflow aggregation with one-sided RDMA (labeled \texttt{ovfl}); (iv) traditional aggregation, flushing at every 4K buffer marks (labeled \texttt{trad}); and (v) for the sake of analysis, \texttt{max-raw} reports the throughput using send-based RDMA in \texttt{DTutils}, \textbf{excluding} any copying costs. The \texttt{send}, \texttt{write}, \texttt{trad}, and \texttt{ovfl} metrics, \textbf{include} the copying costs. Importantly, \texttt{max-raw} in send-based RDMA at the largest buffer sizes (say, 4K) tends to the theoretical maximum using \emph{one-sided RDMA}, because the impact of polling becomes increasingly smaller at larger transfers.
We test three different sizes of remote invocations (small/8B, medium/64B and large/256B). We expect that most applications will perform remote invocations with context and parameters in that size magnitude. Each reported number is an average of three experiments, with each transmitting 65 million messages.

\begin{table}[!htb]
    \centering
    \begin{tabular}{|c|c|c|c|c|c|}
    \hline
        size(B) & send & write & trad & ovfl & max-raw \\ \hline
    \hline
    8 & 12.07 & 38.39 & 2116.78 & 36.52 & 35.84 \\
    64 & 112.95 & 301.47 & 4256.14 & 283.15 & 291.93 \\
    256 & 367.08 & 1129.82 & 5318.89 & 1061.37 & 1135.74 \\ \hline
    \end{tabular}
    \caption{\texttt{call} throughput (MB/s): send-based (\texttt{send}), one-sided (\texttt{write}), one-sided + local aggregation (\texttt{trad}), one-sided + overflow aggregation (\texttt{ovfl}). Control is send-based without serialization (\texttt{max-raw})}.
    \label{tbl:call_performance}
\end{table}

As \texttt{write}/\texttt{ovfl} (one-sided RDMA) closely follow \texttt{max-raw} (send-based RDMA) at their equivalent buffer sizes, we can visualize how one-sided RDMA is much better-performing than send-based RDMA: \texttt{write} and \texttt{ovfl} include the cost of function serialization (with sizes between 8B-256B), but \texttt{max-raw} does not. All metrics for remote invocation (all but \texttt{max-raw}) also include the invocation overhead of an empty function (not elided by the compiler), although we measured no impact in the overall transfer throughput, which is expected. So, with one-sided RDMA, we have low overhead both in function serialization done at the sender, and function invocation done at the receiver. In fact, ``margins are tight'': even the very thin code layer of \texttt{RDMAAggregator} in \texttt{ovfl} mode, which directly forwards calls to an underlying \texttt{RDMAMessenger} as long as no local buffering is performed, imposes a 9\% performance slowdown in throughput.


As discussed in Sec.~\ref{Sec-SendBased}, our one-sided services (\texttt{RDMAAggregator} and \texttt{RDMAMessenger}) use send-based RDMA in their initial setup. \texttt{RDMAAggregator} in \texttt{ovfl} mode is at most 6.2\% under \texttt{max-raw} up to 256B (1065 MB/s vs 1135 MB/s), including function serialization and invocation overhead. The biggest performance gain comes with \texttt{RDMAAggregator} in \texttt{trad} mode (flushing at every 4KB data mark): as function sizes increase, its throughput quickly converges to at most 2.5\% under our theoretical maximum, namely \texttt{max-raw} for 4KB (5318 MB/s in 256B calls in \texttt{trad} vs 5464 MB/s for 4KB transfers in \texttt{DTutils}). The throughput variation of \texttt{RDMAAggregator} in \texttt{trad} mode at different function sizes is indicative of the function serialization overhead, although this overhead is sublinear in relation to decreasing function sizes.



\subsection{Usability and Distributed MCTS}
\label{Sec-Experiments-MCTS}

We evaluate the usability of \sysname by implementing an MCTS \emph{distributed framework}, a higher-level middleware itself that performs the MCTS computation in a distributed fashion given a user-provided MCTS application specification. The following section discusses the reasons why MCTS is considered as appropriate for that goal. The parallelization approach for our MCTS framework is the tree-parallel approach~\cite{TreeParallel}, discussed in Sec.~\ref{Sec-BackgrondRelatedWork}. We test our MCTS framework (and hence the performance and usability aspects of \sysname) with the board game \emph{Hex}.

\sysname consists of around 10050 non-empty C++14-conformant LoC, while the MCTS framework (considered as ``application'' relative to \sysname) contains three components: (a) a class representing tree nodes (475 non-empty, C++14-conformant LoC); (b) the worker thread functions that start rollouts over distributed tree nodes (210 non-empty, C++14-conformant LoC); and (c) the main function, which specifies configuration, and launches/places threads within the system (30 non-empty, C++14 conformant LoC).  Finally, the \emph{Hex} specification (considered as ``application'' now relative to the MCTS distributed framework) contains about 200 non-empty LoC. So, the MCTS distributed framework (MCTS logic) + the MCTS application specification (game logic) consist together only around 8\% of our implemented codebase.

\subsubsection{Rationale for System Validation}
\label{Sec-RationaleExperimentMCTS}

Apart from applicability considerations of MCTS (discussed in Sec.~\ref{Sec-BackgrondRelatedWork}), a distributed MCTS framework is appropriate to evaluate \sysname since: (i) it relies on a distributed data structure, as we choose to employ the tree-parallel distribution strategy of~\cite{TreeParallel}; (ii) it is a memory-bound problem, which suggests that it could highly benefit from the NUMA-awareness in \sysname not-only in the remote invocations themselves, but also in the application-level, as we expose to applications our ``internal'' tools for NUMA-aware, automatic memory management; (iii) depending on the application, MCTS can behave rather irregularly, as the best choices in the MCTS computation at each node might be consistently biased toward certain children. The combination of factors (i) and (iii) is particularly challenging, because while distributing MCTS nodes uniformly at random across machines is applicable to irregularity, it completely destroys the communication locality. In theory, low-overhead, aggregate communication would mitigate these problems substantially, and we certainly intend to evaluate this aspect of \sysname as it reflects in a real-world relevant application.

\subsubsection{Setup and Usability Considerations}

An MCTS tree node is represented by class \texttt{MCTS\_Node}. Each \texttt{MCTS\_Node} is ``owned'' by a particular thread. A global variable \texttt{root} initially points to the unique \texttt{MCTS\_Node} in the tree, owned by thread 0. We denote the number of threads as $n$, the number of processes as $n_p$, and the number of machines $n_m$.

In our experiment, the computation is divided into \emph{phases}. In each phase, all threads residing in the same process as the root owner start rollouts, up to a maximum of $4K \cdot n$ rollouts. When an expansion needs to be performed in a node $N_p$, owned by thread $T_p$, a thread is selected uniformly at random to become the owner $T_c$ of a child node $N_c$. A (possibly remote) \texttt{call\_buffer} is done for $T_c$, which will create the new \texttt{MCTS\_Node} $N_c$, later notifying the parent of the new node location, by means of a \texttt{call}. We create children nodes using \texttt{call\_buffer} as we have to pass a buffer representing the game state associated with $N_p$. When a selection needs to be performed in the (possibly remote) node $N_c$, a \texttt{call} is performed on the appropriate child in order to continue moving down the tree. We can easily set the MCTS framework to only perform remote calls, so even when $T_p \ne T_c$, if they both share a process, $T_p$ will create the child on behalf of $T_c$. We did not see any statistically-significant change in performance under this setting, indicative of low-overhead in remote call serialization. 

Note that since all threads in the same process as the root node operate concurrently on it, there might be a gap between (a) when a \texttt{call} is made for $T_c$ in order to create $N_c$, and (b) when $T_c$ actually finishes notifying the parent of $N_c$'s location. We have a mechanism called \emph{deferred selection}, which essentially accumulates selection requests in the parent, and relies on the child notification to, in addition to notifying the parent $N_p$ of the child $N_c$'s address, resume deferred selections at the parent, directing them to the child. 

When the selection process reaches a leaf node, a \emph{simulation request} is distributed uniformly at random among the threads that reside locally on such a leaf node, using wait-free concurrent queues. Upon consuming from those queues, worker threads perform random simulations (16 by default), and perform backpropagation upwards, using again distributed \texttt{call} and \texttt{call\_buffer}. The selection updates \texttt{visit} counters on the nodes as it moves down the tree, and the backpropagation updates \texttt{win} counters moving upwards. Each node has such counters associated with each of its children, in order to implement selection.

Notably, most code in the MCTS framework layer is related to message synchronization (e.g. deferred selections), or to shared-memory synchronization (e.g wait-free concurrent data structures to store children pointers, relaxed-memory atomic operations to update visit/win counters in the virtual loss technique, etc). In the MCTS framework, the \emph{only} aspects related to inter-thread communication are (a) the creation of \texttt{RDMAAggregator}s, a one-time operation per worker thread when it launches; (b) performing remote calls using \texttt{call} and \texttt{call\_buffer} invocations inside \texttt{MCTS\_Node}; and (c) shutting down the \texttt{RDMAAggregator}s, which sends a unidirectional shutdown signal to the symmetrical \texttt{RDMAAggregator}s, also a one-time operation per worker thread. While \sysname allows users to obtain chunks of remote memory, and copy \emph{data} directly over them, being able to perform \emph{remote invocations} over remote machines proved incredibly convenient, from an implementer's perspective, for the completion of the MCTS framework.

\subsubsection{Performance Considerations}

\begin{figure*}[!htb]
	\centering
	\includegraphics[width=\textwidth]{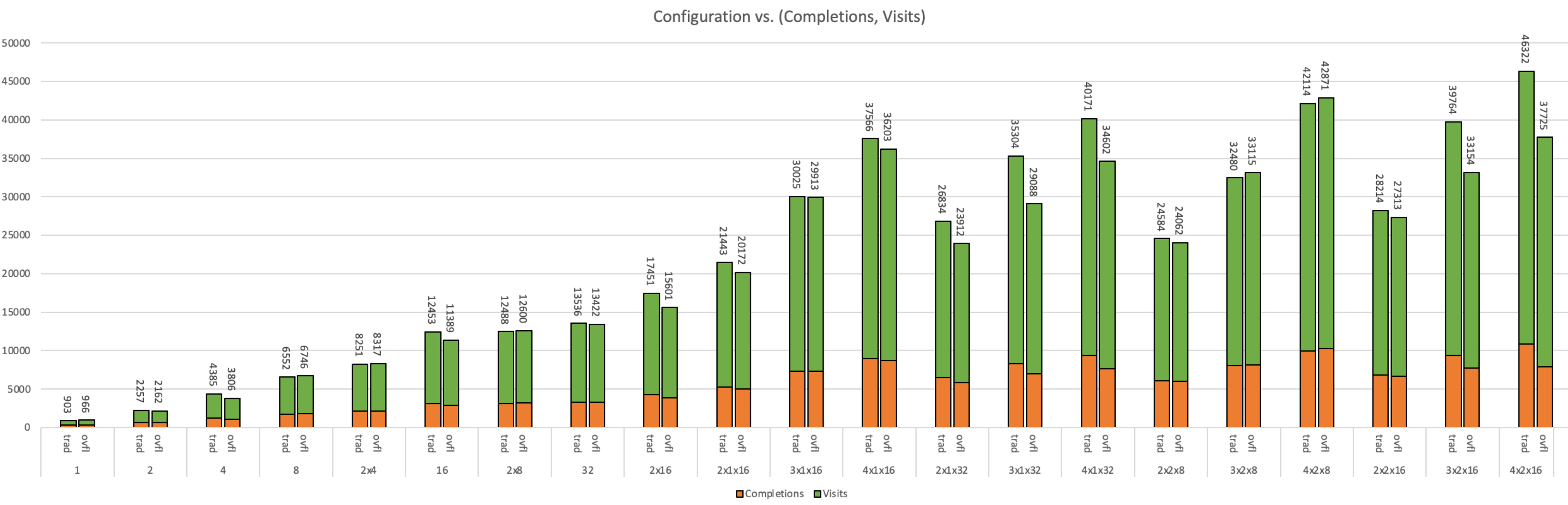}
	\caption{Different configurations ($x$) vs. visits and (completions + visits) ($y$).}
	\label{fig:configuration_performance}
\end{figure*}

We tested many configurations for process/thread placement, and show results in Fig.~\ref{fig:configuration_performance}. In our experiments, threads communicate using \texttt{RDMAAggregator}s, in \texttt{trad} or \texttt{ovfl} mode, as discussed in Sec.~\ref{Sec-RemoteInvocationSubsystem}. In the figure, a configuration in shown in the form $m \times p \times t$, where $m$ is the number of machines, $p$ is the number of processes per machine, and $t$ is the number of threads per process. Prefixes $1 \times$ or $1 \times 1 \times$ are omitted.

We recall that each machine has 2 NUMA zones, each with 8 cores and 16 hyperthreads. Within one machine, single-process configurations 1, 2, 4, and 8 pin threads to cores within a single NUMA zone. Configuration 16 has 8 threads in each NUMA zone, avoiding hyperthreading, and configuration 32 has 16 threads in each NUMA zone, relying on hyperthreading. Within one machine, two-process configurations 2x4, 2x8, and 2x16 each have one process in its own NUMA zone, without hyperthreading for 2x4 and 2x8, and with hyperthreading for 2x16.

In Fig.~\ref{fig:configuration_performance}, we note scalability within a single NUMA zone up to 4 processes, at which point the 2x4-configuration outperforms the 8-configuration. Despite a bigger cost with inter-process remote calls in the 2x4-configuration (rather than only intra-process calls in the 8-configuration), we have less memory pressure on the cache-coherence protocol, because at most 4 threads (rather than 8 threads) concurrently operate on a single memory operation in order to update visits/wins counters as to implement the virtual loss technique. The 2x8- and 2x16-configurations benefit similarly compared to the 16- and 32-configurations, but in \texttt{trad} mode the 2x8-configuration (resp. 2x16-configuration) scales at slope 0.75 compared to the 2x4-configuration (resp. 2x8-configuration), where the ideal would be 1.0. Application profiling seems to indicate that this is due to (i) increased memory pressure on the atomic operations for visits/wins counters; and (ii) the impact of hyperthreading, as relative costs of functions are similar among the configurations. In \texttt{ovfl} mode, there is an additional penalty with high thread counts compared to 16 threads. We measured the maximum number of buffered chunks in \texttt{ovfl} mode, which happens whenever threads generate remote calls faster than receivers consume them. This number was bigger than 1 in only 0.016\% of symmetrical \texttt{RDMAMessenger}s, so threads almost always had chunk space in order to perform remote writes, without waiting for remote allocations. The reason for the overhead in \texttt{ovfl} mode is simply the 20x to 60x worse performance compared to \texttt{trad} mode with remote invocations in the order of 8-64 B in size, as seen in Tbl.~\ref{tbl:call_performance}, in the previous section. With larger thread counts, traditional aggregation pays off. The 2x2x8, 3x2x8, and 4x2x8 represent optimal NUMA placement with less memory pressure as we have two processes per machine, as discussed before, and they represent our best-scaling configuration as we double the number of threads up to 64 physical cores among the machines.

\section{Conclusion}
\label{Sec-Conclusion}

We present \sysname, a middleware that provides integrated RDMA-based remote invocation, NUMA-aware remote memory allocation, asynchronous data transfer, and message aggregation in idiomatic C++1x. As \sysname aims to support distributed data structures and applications sensitive to latency or subject to irregular communication, we evaluate the usability of our system by implementing a higher-level Monte-Carlo Tree Search framework. Our MCTS framework executes a user-provided MCTS problem specification in a distributed fashion, and the MCTS problem specifications do not require users to implement any code related to RDMA communication, memory management for RDMA (handled by \sysname), or MCTS logic (handled by the MCTS framework). We show that our remote invocations have low overhead with a comparative micro-benchmark, and our MCTS application framework, while simulating plays of the board game Hex, scales well up to the number of non-hyperthreaded CPU cores in our testing environment.


\begin{thebibliography}{10}

\bibitem{AdaptiveMultistageSampling}
Peter Auer, Nicolo Cesa-Bianchi, and Paul Fischer.
\newblock Finite-time analysis of the multiarmed bandit problem.
\newblock {\em Machine learning}, 47(2):235--256, 2002.

\bibitem{UpperConfidenceBound}
Peter Auer, Nicolo Cesa-Bianchi, and Paul Fischer.
\newblock Finite-time analysis of the multiarmed bandit problem.
\newblock {\em Machine learning}, 47(2):235--256, 2002.

\bibitem{infinity}
Claude Barthels, Simon Loesing, Gustavo Alonso, and Donald Kossmann.
\newblock Rack-scale in-memory join processing using rdma.
\newblock In {\em Proceedings of the 2015 ACM SIGMOD International Conference
  on Management of Data}, SIGMOD '15, pages 1463--1475, New York, NY, USA,
  2015. Association for Computing Machinery.

\bibitem{rdmaDB2}
Claude Barthels, Simon Loesing, Gustavo Alonso, and Donald Kossmann.
\newblock Rack-scale in-memory join processing using rdma.
\newblock In {\em Proceedings of the 2015 ACM SIGMOD International Conference
  on Management of Data}, SIGMOD '15, pages 1463--1475, New York, NY, USA,
  2015. Association for Computing Machinery.

\bibitem{rdma-reg2}
C.~Bell and D.~Bonachea.
\newblock A new dma registration strategy for pinning-based high performance
  networks.
\newblock In {\em Proceedings International Parallel and Distributed Processing
  Symposium}, pages 10 pp.--, 2003.

\bibitem{delegation}
Irina Calciu, Dave Dice, Tim Harris, Maurice Herlihy, Alex Kogan, Virendra
  Marathe, and Mark Moir.
\newblock Message passing or shared memory: Evaluating the delegation
  abstraction for multicores.
\newblock In Roberto Baldoni, Nicolas Nisse, and Maarten van Steen, editors,
  {\em Principles of Distributed Systems}, pages 83--97. Springer International
  Publishing, 2013.

\bibitem{RootLeafParallel}
Tristan Cazenave and Nicolas Jouandeau.
\newblock On the parallelization of uct.
\newblock In {\em Computer games workshop}, 2007.

\bibitem{TreeParallel}
Guillaume MJ-B Chaslot, Mark~HM Winands, and H~Jaap van Den~Herik.
\newblock Parallel monte-carlo tree search.
\newblock In {\em International Conference on Computers and Games}, pages
  60--71. Springer, 2008.

\bibitem{ib-latency}
NVIDIA Corporation.
\newblock www.nvidia.com/en-us/networking/products/infiniband.

\bibitem{rdmaFarm}
Aleksandar Dragojevic, Dushyanth Narayanan, Miguel Castro, and Orion Hodson.
\newblock Farm: Fast remote memory.
\newblock In {\em 11th USENIX Symposium on Networked Systems Design and
  Implementation (NSDI 2014)}. USENIX -- Advanced Computing Systems
  Association, April 2014.

\bibitem{combining}
Panagiota Fatourou and Nikolaos~D. Kallimanis.
\newblock Revisiting the combining synchronization technique.
\newblock In {\em Proceedings of the 17th ACM SIGPLAN Symposium on Principles
  and Practice of Parallel Programming}, PPoPP '12, pages 257--266, New York,
  NY, USA, 2012. ACM.

\bibitem{rdma-reg}
Balazs Gerofi, Masamichi Takagi, and Yutaka Ishikawa.
\newblock Revisiting rdma buffer registration in the context of lightweight
  multi-kernels.
\newblock In {\em Proceedings of the 23rd European MPI Users' Group Meeting},
  EuroMPI 2016, pages 180--183, New York, NY, USA, 2016. Association for
  Computing Machinery.

\bibitem{TopologyHeuristic}
Alonso Gragera and Vorapong Suppakitpaisarn.
\newblock A mapping heuristic for minimizing message latency in massively
  distributed mcts.
\newblock In {\em 2016 Fourth International Symposium on Computing and
  Networking (CANDAR)}, pages 547--553. IEEE, 2016.

\bibitem{Hex}
Ryan~B Hayward and Bjarne Toft.
\newblock {\em Hex: The Full Story}.
\newblock CRC Press, 2019.

\bibitem{progress}
Maurice Herlihy and Nir Shavit.
\newblock On the nature of progress.
\newblock In Antonio Fern{\`a}ndez~Anta, Giuseppe Lipari, and Matthieu Roy,
  editors, {\em Principles of Distributed Systems}, pages 313--328, Berlin,
  Heidelberg, 2011. Springer Berlin Heidelberg.

\bibitem{rdmaKV1}
Anuj Kalia, Michael Kaminsky, and David~G. Andersen.
\newblock Using rdma efficiently for key-value services.
\newblock {\em SIGCOMM Comput. Commun. Rev.}, 44(4):295--306, August 2014.

\bibitem{beyondMPI}
Feilong Liu, Claude Barthels, Spyros Blanas, Hideaki Kimura, and Garret Swart.
\newblock Beyond mpi: New communication interfaces for database systems and
  data-intensive applications.
\newblock {\em SIGMOD Rec.}, 49(4):12--17, March 2021.

\bibitem{rdmaKV2}
Christopher Mitchell, Yifeng Geng, and Jinyang Li.
\newblock Using one-sided rdma reads to build a fast, cpu-efficient key-value
  store.
\newblock In {\em Proceedings of the 2013 USENIX Conference on Annual Technical
  Conference}, USENIX ATC'13, pages 103--114, USA, 2013. USENIX Association.

\bibitem{grappa}
Jacob Nelson, Brandon Holt, Brandon Myers, Preston Briggs, Luis Ceze, Simon
  Kahan, and Mark Oskin.
\newblock Latency-tolerant software distributed shared memory.
\newblock In {\em 2015 USENIX Annual Technical Conference (USENIX ATC 15)},
  pages 291--305, Santa Clara, CA, 2015. USENIX Association.

\bibitem{rdmaNUMA}
Jacob Nelson and Roberto Palmieri.
\newblock Performance evaluation of the impact of {NUMA} on one-sided {RDMA}
  interactions.
\newblock In {\em International Symposium on Reliable Distributed Systems,
  {SRDS} 2020, Shanghai, China, September 21-24, 2020}, pages 288--298. {IEEE},
  2020.

\bibitem{rdmaStorm}
Stanko Novakovic, Yizhou Shan, Aasheesh Kolli, Michael Cui, Yiying Zhang,
  Haggai Eran, Boris Pismenny, Liran Liss, Michael Wei, Dan Tsafrir, and Marcos
  Aguilera.
\newblock Storm: a fast transactional dataplane for remote data structures.
\newblock In {\em 12th ACM International Systems and Storage Conference
  (SYSTOR)}. ACM, USENIX, June 2019.
\newblock Best Paper Award.

\bibitem{AlphaGo}
David Silver, Aja Huang, Chris~J Maddison, Arthur Guez, Laurent Sifre, George
  Van Den~Driessche, Julian Schrittwieser, Ioannis Antonoglou, Veda
  Panneershelvam, Marc Lanctot, et~al.
\newblock Mastering the game of go with deep neural networks and tree search.
\newblock {\em nature}, 529(7587):484--489, 2016.

\bibitem{RootTreeParallelization}
Maciej {\'S}wiechowski and Jacek Ma{\'n}dziuk.
\newblock A hybrid approach to parallelization of monte carlo tree search in
  general game playing.
\newblock In {\em Challenging Problems and Solutions in Intelligent Systems},
  pages 199--215. Springer, 2016.

\bibitem{rdmaLITE}
Shin-Yeh Tsai and Yiying Zhang.
\newblock Lite kernel rdma support for datacenter applications.
\newblock In {\em Proceedings of the 26th Symposium on Operating Systems
  Principles}, SOSP '17, pages 306--324, New York, NY, USA, 2017. Association
  for Computing Machinery.

\bibitem{ZobristHashing}
Kazuki Yoshizoe, Akihiro Kishimoto, Tomoyuki Kaneko, Haruhiro Yoshimoto, and
  Yutaka Ishikawa.
\newblock Scalable distributed monte-carlo tree search.
\newblock In {\em SoCS}, pages 180--187, 2011.

\bibitem{rdmaDB1}
Erfan Zamanian, Carsten Binnig, Tim Harris, and Tim Kraska.
\newblock The end of a myth: Distributed transactions can scale.
\newblock {\em Proc. VLDB Endow.}, 10(6):685--696, February 2017.

\end{thebibliography}

\newpage 

\appendix

\section{Real Code Example}
\label{Sec-RealCodeExample}

We provide a real C++11 example of \sysname's features on Lst.~\ref{Alg-Demo} below. Line 1 initializes the thread-local allocators discussed Sec.~\ref{Sec-NUMA-MemoryRegistration}), and by default we allocate one per thread. Line 2 initializes one \texttt{RDMAMessenger} per destination thread (see Sec.~\ref{Sec-RemoteInvocationSubsystem}) as we instantiate one \texttt{RDMAMessengerGlobal}. Line 6 allocates registered memory in the thread-local circular buffer, on which we produce data in line 7. Line 8 performs a remote \texttt{call\_buffer} on the \texttt{RDMAMessenger} associated with the destination thread, using one-sided RDMA. Note the similarity with line 21, where we perform an invocation using send-based RDMA. In line 17, the thread handles messages received from others, noting that handling remote invocations could be delegated to separate helper threads. Naturally, in this case, the application is expected to make sure that remote calls use proper synchronization to avoid data races with respect to the destination thread and to other helper threads while accessing shared objects.

\begin{lstlisting}[float=*, numbers=left, stepnumber=3,
	caption={\textbf{Real} C++ code demo. Please refer to additional discussion in the text.},
	label=Alg-Demo]
void demonstration_thread(int offset) {
    seriema::init_thread(offset);
    RDMAMessengerGlobal messenger;

    for(uint64_t i = 0; i < number_operations; i++) {
        RDMAMemory *registered_data = seriema::thread_information->circular_allocator->allocate(4096);
        int filled_size = fill_data(registered_data->get_buffer()); // User function
        bool result = messenger.call_buffer<RetryAsync>(destination_thread,
            [source_thread = i](void *buffer, uint64_t size) {
                // Remote function body:
                //  - can use "source_thread", "buffer", "size"
                //  - "buffer" has all "size" bytes filled in the source thread
                //  - requires synchronization only if it accesses concurrent objects
            },
            source, 0, filled_size, synchronizer);
        // Processing remote messages can also be delegated to helper threads
        if(i % 128 == 0) { messenger.process_calls_all(); }
    }

    Synchronizer remote_synchronizer;
    seriema::call<RemoteNotify>(0, [source_thread = i] {
        // Remote function body:
        //  - "source_thread" notified it is done
    }, remote_synchronizer);
    remote_synchronizer.spin_nonzero_operations_left();

    messenger.shutdown_all();
    seriema::finalize_thread();
}
\end{lstlisting}

Lines 18 and 21 also demonstrate how lambda captures on the source thread are serialized and become available for the destination threads in a natural way. Finally, every \texttt{call} or \texttt{call\_buffer} invocation, when using one-sided RDMA, can specify what to do in case the remote buffer where we write serialized calls becomes full. If \texttt{Retry} is specified, the invocation blocks until remote buffers become available, and if \texttt{RetryAsync} is specified, the serialized invocation is pushed to a buffer where helper threads asynchronously try to finish the operation, guaranteeing a per-thread FIFO order. On line 21, the \texttt{RemoteNotify} specifies that once the receiver consumed the call, it will notify the \texttt{Synchronizer} in the source, in this case with send-based RDMA. If the \texttt{Synchronizer} passed to the remote invocation was wrapped inside a \texttt{RemoteObject} class, the notification for the sender would be done by one-sided RDMA. The sender blocks until the notification is received in line 25.

\end{document}